\documentclass[twocolumn,english,aps,prl,showpacs,longbibliography,showkeys,notitlepage]{revtex4-1}
\usepackage[T1]{fontenc}
\usepackage{lmodern}
\usepackage{color}
\usepackage{babel}
\usepackage{amsmath}
\usepackage{amssymb}
\usepackage{graphicx}

\RequirePackage[colorlinks=true,linkcolor=blue]{hyperref}
\usepackage{dsfont}
\usepackage{cleveref}
\usepackage{natbib}
\usepackage{bm}
\usepackage{blindtext}
\usepackage{pdfpages}

\usepackage{amsfonts}
\usepackage{babel}
\usepackage{braket}
\usepackage{verbatim}
\usepackage{amstext}
  
\makeatletter
\AtBeginDocument{\let\LS@rot\@undefined}
\makeatother

\begin{document}

\title{Hybrid superconductor-semiconductor systems for quantum technology}

\author{M. Benito}
\address{Department of Physics, University of Konstanz, D-78457 Konstanz, Germany}
\author{Guido Burkard}
\address{Department of Physics, University of Konstanz, D-78457
  Konstanz, Germany}

\begin{abstract}
Superconducting quantum devices provide excellent connectivity and controllability while semiconductor spin qubits stand out with their long-lasting quantum coherence, fast control, and potential for miniaturization and scaling. In the last few years, remarkable progress has been made in combining superconducting circuits and semiconducting devices into hybrid quantum systems that benefit from the physical properties of both constituents. Superconducting cavities can mediate quantum-coherent coupling over long distances between electronic degrees of freedom such as the spin of individual electrons on a semiconductor chip and thus provide essential connectivity for a quantum device.  Electron spins in semiconductor quantum dots have reached very long coherence times and allow for fast quantum gate operations with increasing fidelities.  We summarize recent progress and theoretical models that describe superconducting-semiconducting hybrid quantum systems,
explain the limitations of these systems, and describe different directions where future experiments and theory are headed.
\end{abstract}

\maketitle

%
Quantum dots (QDs) are nanostructures hosted in semiconductors where a few electrons can be electrostatically trapped in discrete states. Therefore, QDs allow to access and control the quantum nature of single electrons and interactions between them~\cite{Hanson2007}. 
Since the first measurements of few-electron phenomena  in lateral gated QDs, the focus of applications of these systems has shifted from  single spintronics  towards quantum information science, as originally envisioned by Loss and  DiVincenzo~\cite{Loss1998}.
Beyond this first proposal, which employs the electron spin as a quantum bit (qubit), the field has developed both theoretically and experimentally as gate and memory fidelities have increased and more complex but robust alternative implementations of spin qubits have been demonstrated, such as a spin qubit defined with a pair of electrons in two QDs~\cite{Levy2002} (singlet-triplet qubit), and  three QDs filled with one, two and three electrons~\cite{Kyriakidis2007,Shi2012,DiVincenzo2000}.

The demonstrated long spin coherence times of electrons in silicon~\cite{Zwanenburg2013} have motivated a change of trend from the traditional host material GaAs to silicon-based QDs. Among the advantages, silicon offers an almost nuclear-spin-free environment for the electronic spins and a significantly reduced spin-orbit interaction, main sources of decoherence in GaAs QDs~\cite{Hanson2007,Kloeffel2013,Awschalom2013}.  
An impressive series of promising quantum information experiments have been realized with silicon spin qubits, including high fidelity single-qubit gates~\cite{Veldhorst2014,Takeda2016,Yoneda2018,Yang2019}, two-qubit gates~\cite{Veldhorst2015, Zajac2017,Watson2018,Huang2018,Xue2018}  and quantum non-demolition measurement~\cite{Xue2019,Yoneda2019}, but scaling these QD systems  to large numbers of qubits  is still challenging, due to the large number of voltage gates and the lack of connectivity due to the intrinsically short-range exchange interaction.
Moreover, since silicon is an indirect bandgap semiconductor, it comes with the additional--and often obstructive-- valley degree of freedom, which is not yet well understood given the complexity of the heterostructures.
Although there are some measurements and estimations of valley features including the valley splitting~\cite{Borselli2011,Zajac2015}, for scalable spin qubit platforms based on silicon QDs~\cite{Zajac2016,Veldhorst2017b,Vandersypen2017} an accurate characterization of all valley features that could affect the fidelity of the computation is desirable.

\begin{figure}
\includegraphics[width=1\columnwidth]{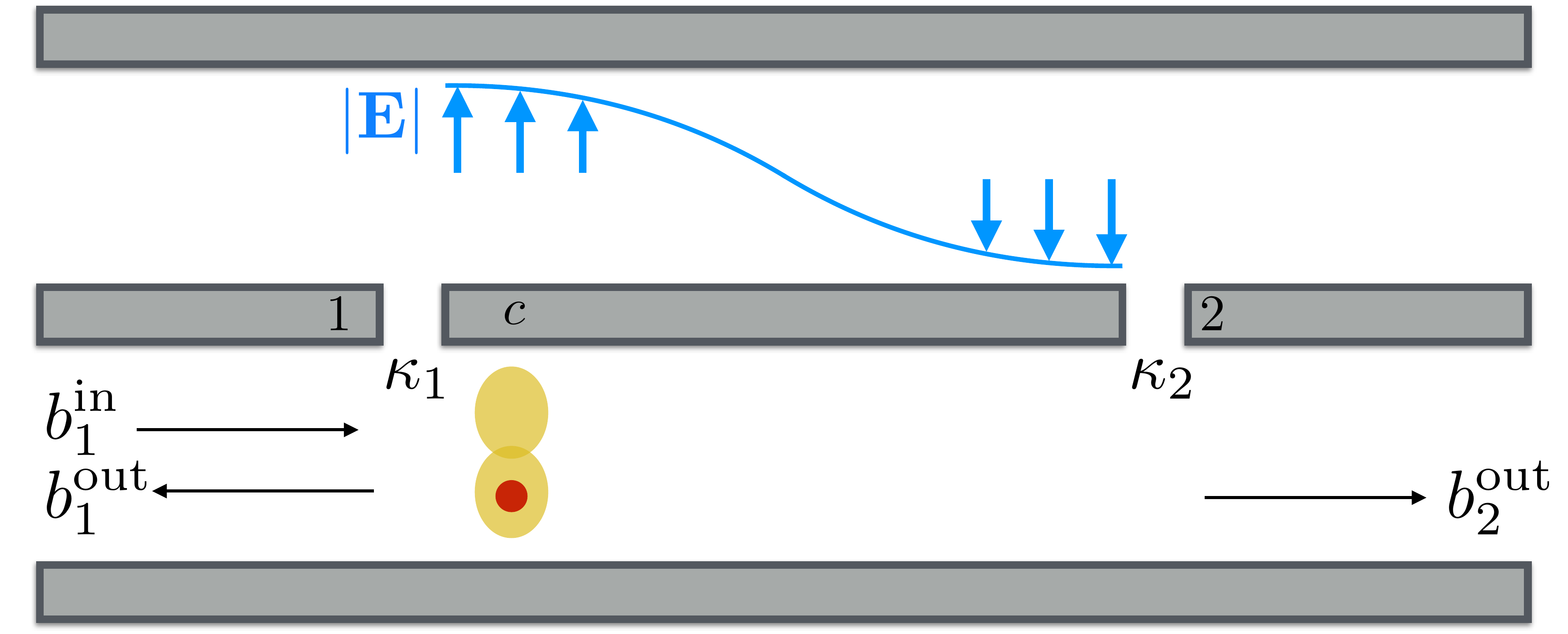}
\protect{\caption{\label{fig:circuitQED}  Schematic representation of a  superconductor-semiconductor hybrid system. 
The superconducting microwave cavity (in grey) can be probed with microwave fields $b_1$ ($b_2$) from port 1 (2) which is coupled to the center conductor (c) with a coupling rate $\kappa_1$ ($\kappa_2$).
A single electron (red dot) in the embedded semiconductor double QD (yellow) interacts with the cavity electric field ${\bf E}$ (blue curve) via the electric dipole coupling.}}
\end{figure}

Impressive progress towards overcoming these challenges  is occurring thanks to the development of superconductor-semiconductor hybrid systems~\cite{Childress2004,Burkard2006,Delbecq2011,Frey2012,2017arXiv170900466C,Burkard2019,Petersson2012,Mi2017,Mi2017b,Stockklauser2017,Bruhat2018,Woerkom2018,Koski2019}, where semiconductor QDs are coherently coupled to superconducting cavities; see Fig.~\ref{fig:circuitQED}.
Hybrid systems mimic  atomic cavity quantum electrodynamics (QED) systems, in which coherent interactions and quantum superpositions of light and matter were successfully demonstrated~\cite{Haroche1989,Raimond2001,Haroche2006}. In fact, the hybrid systems of interest here are more similar to the so-called circuit QED systems where superconducting qubits are coupled to superconducting cavities~\cite{Wallraff2004,Blais2004}. 
An effective implementation of the cavity photons as mediators would allow one to simplify the QD qubit architecture and increase its connectivity.
%

An ensemble of spins can be resonantly coupled to the field in a superconducting cavity via its large effective magnetic dipole~\cite{Imamoglu2009,Schuster2010,Amsuss2011}, but the  coupling of the tiny magnetic dipole of a single electron to a cavity remains difficult.
Different mechanisms and techniques to couple single-electron spin qubits~\cite{Childress2004,Trif2008,Cottet2010,Hu2012} and multi-electron spin qubits~\cite{Burkard2006,Jin2012,Guido_RX_PRB_2015,Srinivasa2016} to superconducting cavities have been theoretically proposed.  However, despite their differences, all of these mechanisms imply using electronic systems with more than the two quantum levels required for the qubit itself, and thereby endowing the electron spin with an effective electric dipole that interacts strongly with the cavity electric field.
First experimental observations of signatures of single-spin to cavity coupling \cite{Petersson2012,Viennot2015}, were recently followed by demonstrations of spin-photon coupling in the strong coupling regime with single-spin~\cite{Mi2018,Samkharadze2018} and three-spin~\cite{Landig2018} qubits. 
In the strong coupling regime of cavity QED, the matter-light coupling $g$ exceeds both the photon loss rate $\kappa$ from the cavity and the decoherence rate $\gamma$ in the two-level matter system (atom or qubit). This means that energy can be exchanged multiple times in quantum-coherent fashion between light and matter, which is typically a prerequisite for the use of a cavity as a long-distance mediator for quantum information.
Moreover, these mature superconductor-semiconductor hybrid systems recently opened a new way to measure QD valley features~\cite{Burkard2016,Mi2017_PRL}.
 
\begin{figure}
\includegraphics[width=1\columnwidth]{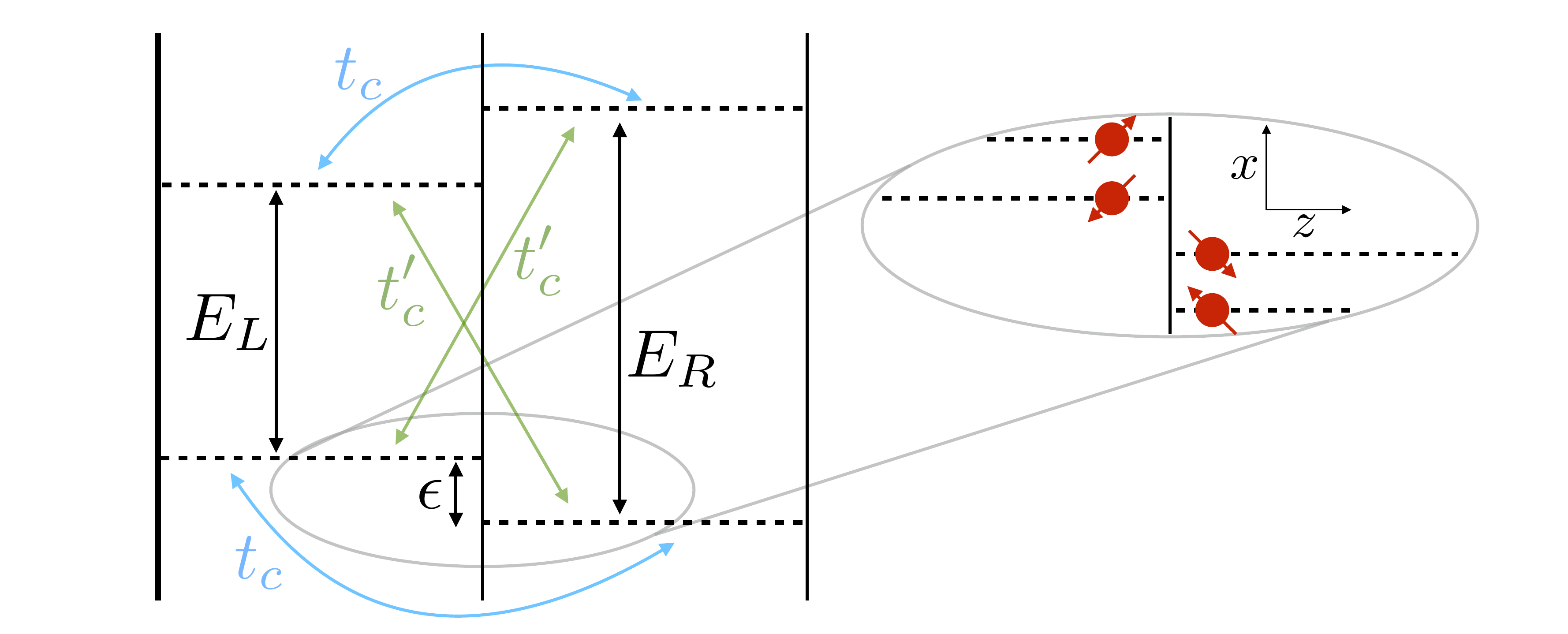}
\caption{\label{fig:DQD} Low-energy levels of a silicon double QD nanostructure with position (L,R), spin ($\uparrow$,$\downarrow$) and valley degrees of freedom. Here, $E_{L(R)}$ denotes the left (right) QD valley splitting, $\epsilon$  the detuning between the left and right QD ground state energy, and $t_c$ ($t_c'$) the intra(inter)-valley tunnel coupling. 
Inside the gray oval, more detail on the spin sublevels of the lower valley states is given. A micromagnet can induce a different magnetic field direction at the center of each QD leading to canted spin quantization axes in the two QDs.}
\end{figure}

 In the following we briefly introduce the theory developed to predict signatures of  the interaction between superconducting cavities and multi-level electronic systems in the cavity transmission, which is a generalized type of input-output theory~\cite{Gardiner1985,Burkard2016,Benito2017,Kohler2018}. Then, we show how this theory can be applied to  the interface between a single electron spin and cavity photons~\cite{Benito2017,Mi2018} and to high-resolution valley spectroscopy~\cite{Burkard2016,Mi2017_PRL,Russ2020}.
 Figure~\ref{fig:DQD} represents a general silicon double QD nanostructure with position, spin and valley degrees of freedom.
 The electrostatic detuning $\epsilon$ between the left and right QDs and the intra(inter)-valley tunnel coupling $t_c$ ($t_c'$) can be controlled externally. One can also apply an external magnetic field and use micromagnets to introduce (static) magnetic field  gradients along the double QD axis.
 For the purpose of spin-photon coupling, it is desirable for the valley splittings  $E_{L,R}$ to exceed the molecular (charge qubit) level splitting $\sqrt{\epsilon^2+4t_c^2}$ as well as the Zeeman splitting due to the external magnetic field $B_z$. Then, a gradient of the transverse magnetic field component  along the double QD axis (z)(such that the magnetic field at the center of the left and right QDs is $B_z \hat{z}\pm b_x \hat{x}$) will induce spin-charge mixing and an effective interaction between the spin and the cavity electric field.
 For valley spectroscopy the valley and charge qubit splittings should be comparable, with the spin states  either degenerate or strongly detuned.

We briefly review a theoretical framework that has allowed the quantitative prediction of the electromagnetic response of a superconducting cavity coupled to an embedded electronic semiconductor system.
The response of a QD-cavity system to a weak microwave probe can be determined using input-output theory~\cite{Collett1984, Gardiner1985}, a treatment that enables the calculation of the fields ($b^{\rm out}_{1}$ and $b^{\rm out}_{2}$ in Fig.~\ref{fig:circuitQED}) emerging from the cavity ports, given the incoming fields ($b^{\rm in}_{1}$; while $b^{\rm in}_{2}$ is possible but will not be considered here). 
Given a cavity frequency $\omega_c$ and a general system-cavity interaction Hamiltonian, $H_{\rm I}=g_c d (a+a^{\dagger})$, mediated by the electric dipole operator $d=\sum_{n,m} d_{n,m}|n\rangle\langle m|$ which describes transitions between QD eigenstates $|n\rangle$ and $|m\rangle$ with $H_{\rm QD}|n\rangle=E_n|n\rangle$~\cite{Petersson2012,Burkard2016},
the quantum Langevin equation for the cavity operator $a$ reads
\begin{equation}
\dot{a}=-i\omega_c a -\frac{\kappa}{2}a+\sqrt{\kappa_{1}} b^{\rm in}_{1}-i g_c d \, ,
\end{equation}
where $\kappa_i$ are the cavity decay rates at the two ports, and
$\kappa=\kappa_1+\kappa_2+\kappa_{\rm int}$ the total photon loss rate, 
with $\kappa_{\rm int}$ the intrinsic losses not related to the cavity ports.

In the weak driving regime we can assume that the electronic system remains near the thermal state, as it would in the absence of any QD-cavity coupling, $g_c=0$, such that 
$p_n$ is the equilibrium population of state $|n\rangle$, which may be given by a Boltzmann distribution, $p_n=e^{-E_n/k_B T}/(\sum_n e^{-E_n/k_B T}) $, or by the solution of the corresponding rate equations in the case of a transport setup~\cite{Mi2017_PRL}.
The evolution of the expectation value of the 
operators $\sigma_{nm}=|n\rangle\langle m|$
to first order in the coupling $g_c$ reads
\begin{equation}
\begin{split}
\langle\dot{\sigma}_{nm}\rangle=&-i(E_m-E_n)\langle\sigma_{nm}\rangle-\sum_{n'm'}\gamma_{nm,n'm'}\langle\sigma_{n'm'}\rangle\\
&-ig_c(\langle a\rangle+\langle a^{\dagger}\rangle)d_{mn}(p_n-p_m) \ .
\end{split} 
\end{equation}
Here we have introduced decoherence processes via the matrix elements $\gamma_{nm,n'm'}$.
In frequency space, this constitutes in general a system of  coupled linear equations for the susceptibilities, defined as $\langle\sigma_{nm}(\omega)\rangle=\chi_{nm}(\omega)\left[\langle a(\omega)\rangle+\langle a^{\dagger}(-\omega)\rangle\right]$~\cite{Benito2017,Kohler2018}. 
In the simplest case of an electron in an aligned double QD ($\epsilon=0$) subject to relaxation and pure dephasing, the susceptibilities read
\begin{equation}
\chi_{10(01)}(\omega)=\frac{g_c(p_0-p_1)}{2t_c\mp\omega\mp i \gamma_c}\ ,
\end{equation}
where $|0\rangle$ is the ground state, $|1\rangle$ the  excited state, and $\gamma_c$ is the total decoherence rate.

Finally, for a cavity with high quality factor probed close to resonance and for a sufficiently small coupling $\left\{\kappa,|\omega-\omega_c|,g_c\sum_{n,m}d_{nm}\chi_{nm}(\omega)\ll\omega_c\right\}$~\cite{Kohler2018}, the cavity transmission 
reads
\begin{equation}
A(\omega)=\frac{\langle b^{\rm out}_2\rangle}{\langle b^{\rm in}_1\rangle}=\frac{-i \sqrt{\kappa_1\kappa_2}}{\omega_c-\omega-i\kappa/2+g_c\sum_{nm}d_{nm}\chi_{nm}(\omega)} \, .
\end{equation}
Depending on the level structure and the driving frequency, sometimes it is useful  to simplify this expression accounting only for the transitions that contribute most to the response~\cite{Burkard2016, Benito2017}.

The input-output theory has been generalized to periodically driven systems~\cite{Kohler2017,Kohler2018} and to more complex systems with vibrational degrees of freedom~\cite{Reitz2019,Liu2019}.

\begin{figure}
\includegraphics[width=1\columnwidth]{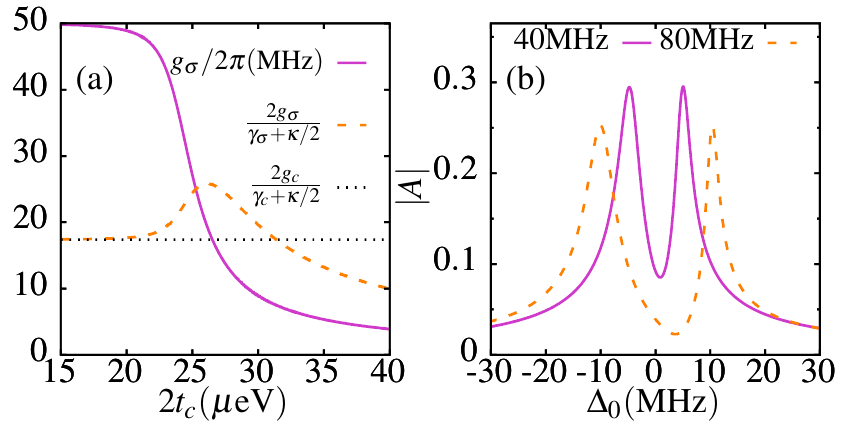}

\protect\caption{\label{fig:newfigure2} 
(a)
Coupling strength $g_{\sigma}$ (solid purple line) as a function of the tunnel splitting $2 t_c$ for a fixed magnetic field profile; $g\mu_B B_z=24\,\mu\text{eV}$ and $g\mu_B b_x=\pm2\,\mu\text{eV}$. 
 Also shown are the ratio between the coupling and decoherence for the spin (dashed orange line) and charge (dotted black line) qubit, for  $g_c/2\pi=50\,\text{MHz}$, $\gamma_c/2\pi=5\,\text{MHz}$, and $\kappa/2\pi=1.5\,\text{MHz}$.
(b)  Vacuum Rabi splitting peaks in the cavity transmission $A$ as a function of detuning $\Delta_0=\omega-\omega_c$ indicating strong spin-photon coupling. The two lines correspond to different values of the charge-photon coupling $g_c/2\pi=\{40,80\}\rm MHz$.
}
\end{figure}

%
The strong coupling regime  for the interaction between a qubit and a cavity implies a coupling rate $g$ larger than the decoherence rates of both the qubit ($\gamma$) and the photon ($\kappa/2$).
The  experimental achievement of strong coupling ($g\gg \gamma,\kappa/2$) between a spin qubit in a QD and a superconducting cavity~\cite{Mi2018,Samkharadze2018,Landig2018,Cubaynes2019}  confirmed that the concept of a spin-based quantum computer with photon-mediated  interactions is feasible. The demonstrations in~\cite{Mi2018,Samkharadze2018,Cubaynes2019} treat a single electron spin qubit delocalized within a double QD and under the influence of a magnetic field gradient perpendicular to the  main quantization axis of the spin, while the demonstration~\cite{Landig2018} does not require magnetic fields but employs an exchange-only (three-electron) spin qubit in a triple QD.  In both cases the spin qubit acquires an electric dipole moment that interacts with the cavity electric field.
In the single electron case,
the strength of this dipole coupling can be tuned by controlling the energy difference between the charge qubit tunnel splitting $2t_c$ and the magnetic Zeeman splitting $g\mu_B B$, and allows for a compromise between  a charge qubit with a short coherence that couples strongly to cavity photons and a much protected pure spin qubit with negligible coupling to cavity photons.
The coupling strength of the low-energy  spin qubit to the cavity can be calculated exactly by a simple diagonalization of the double QD electronic Hamiltonian and, for a symmetric double QD ($\epsilon=0$), it reads $g_{\sigma}=g_c\sin{(\phi_+/2+\phi_-/2)}$, with the spin-charge mixing angles $\phi_{\pm}=\arctan{\left[g\mu_B b_x/(2t_c\pm g\mu_B B_z)\right]}\in [0,\pi)$~\cite{Benito2019}.
It turns out that there is an optimal point for coherent spin-photon coupling  in terms of the relation between the interdot tunnel coupling and the externally applied magnetic field \cite{Beaudoin2016,Benito2017}.
This is summarized in Fig.~\ref{fig:newfigure2}(a), where we show the coupling strength $g_{\sigma}$ of the low-energy spin qubit to the cavity as a function of $t_c$ (for  $\epsilon=0$) and the ratio between the coupling strength $g_\sigma$ and the total decoherence rate $\gamma_\sigma+\kappa/2$, which has a maximum at the optimal point~\cite{Benito2019}. 
The dotted black line corresponds to the same ratio for a charge qubit.
This comparison is valid under the assumption that the decoherence rate of the spin qubit $\gamma_{\sigma}$ is dominated by the effects of the hybridization with charge,  where $\gamma_c=1/T_{2c}$ is the total charge  decoherence rate, the inverse of the charge decoherence time $T_{2c}$. 
  In the middle region around $2 t_c \approx 25\,\mu{\rm eV}$
  the advantage obtained by
 using the electron spin with its long coherence time overcompensates the
 concomitant loss in spin coherence due to spin-charge hybridization.
 
 Spin-cavity interaction can be probed by tuning the spin qubit into resonance with the cavity mode, injecting a microwave tone into the cavity, and observing its transmission coefficient.
 In Fig.~\ref{fig:newfigure2}(b) we show the cavity transmission coefficient $A$ as calculated using input-output theory, as a function of the detuning $\Delta_0$ between the probe and cavity frequencies, predicting a well-resolved vacuum Rabi splitting of the cavity resonance peak that was reported in Refs.~\cite{Mi2018,Samkharadze2018}, hallmarking the strong coupling between a single electron spin qubit and a cavity photon.
 The asymmetry between the two peaks, more apparent as $g_c$ increases, is due to the presence of a third energy level and the interplay between the contributions $\chi_{01}$ and $\chi_{02}$~\cite{Benito2017}, where $|0\rangle$ is the ground state and $|1\rangle$ and $|2\rangle$ the first and second excited states.
 
 While pioneering works harnessed the magnetic field gradient generated by a micromagnet to electrically drive spin rotations on an electron spin situated in a single QD~\cite{Pioro-Ladriere2008,Kawakami2014,Yoneda2018}, recent studies have demonstrated
that a double QD configuration with aligned energy levels ($\epsilon=0$) allows for low-power electric dipole spin transitions~\cite{Croot2019} because in this "flopping mode", the electron samples a larger magnetic field range and has a larger electric dipole.
Also in this mode of operation, the cavity-assisted spin readout has been theoretically optimized, with fidelities in the range $80-95\%$ in a few $\mu\text{s}$ being within reach \cite{Danjou2019}.
Moreover, this configuration provides the spin qubit with ``sweet spots'', i.e., points in the parameter space where the qubit is naturally protected from charge detuning fluctuations~\cite{Vion2002,Petersson2010,Kim2014,Benito2019b}.

The coupling to superconducting cavities has also provided high-fidelity readout of a two-electron spin state in a double QD~\cite{Zheng2019}. The so called singlet-triplet qubit, defined with two electrons in a double QD, can be operated in different regimes such that the nature of the coupling to the resonator can change from a standard transverse coupling~\cite{Burkard2006,Jin2012} to a longitudinal coupling~\cite{Ruskov2019,Harvey2018}. Although the strong-coupling regime to a cavity photon has not yet been  demonstrated, recent theory progress in identifying sweet spots~\cite{AbadilloUriel2019}, together with ongoing work to improve experimental devices, should make this possible.

The experiments and  theory discussed so far rely on large valley splittings, such that the valley degree of freedom barely affects the spin dynamics. However, it is worth mentioning that within a QD nanostructure the nature of the low lying valley states, e.g., in silicon-based systems, may change from one QD to the other~\cite{Burkard2016}, which results in inter-valley tunnel coupling and therefore the possibility to define a valley-orbit qubit that has been proven to couple strongly to the cavity photons~\cite{Mi2018b}.
It was theoretically predicted \cite{Burkard2016} and experimentally confirmed \cite{Mi2017_PRL} that the low lying valley features, not only valley splittings but also intra- and inter-valley tunnel couplings, of a few electron silicon QD systems are accessible in a hybrid circuit QED system, since they generate a fingerprint on the cavity transmission. 
This complete information on valley features without the need of a magnetic field make this scheme attractive in comparison to conventional magnetospectroscopic approaches~\cite{Borselli2011,Yang2013}.

What are the prospects of hybrid superconductor-semiconductor systems for quantum technology?
Once the strong coupling between different spin qubits to superconducting cavities had been  demonstrated, another important challenge emerged: to demonstrate spin qubit interactions mediated by cavity photons, in a similar way as previously achieved for superconducting qubits \cite{Majer2007,Fink2009} and double QD charge qubits \cite{Delbecq2013,Woerkom2018,Wang2020}. An important milestone in this direction was to tune  two spin qubits simultaneously into resonance with the cavity and observe a collectively-enhanced splitting in a transmission experiment~\cite{Borjans2019}.
 
The advantage of using the spin rather than the charge is twofold: (i) the spin-cavity coupling can be turned off by increasing the tunnel coupling $t_c$, therefore maintaining the qubit in a sweet spot protected from charge noise, and (ii) the spin-qubit approach holds the potential of reducing the spin-charge mixing, with the corresponding reduction on spin-photon coupling $g_{\sigma}$ and  spin qubit linewidth $\gamma_{\sigma}$, such that eventually the condition $\gamma_{\sigma} \ll \kappa/2$, for which cavity-mediated two-qubit gates and readout fidelities are maximized for the device, is fulfilled~\cite{Benito2019}. 
This optimization  demands a relatively small degree of spin-charge mixing, in order  to make the spin decoherence rate comparable to the Purcell relaxation rate. Therefore, experiments that attempt to demonstrate this effect would greatly benefit from the use of isotopically purified silicon~\cite{Zwanenburg2013}. 

 Recent theory work concludes that  two-qubit gates mediated by cavity photons are capable of reaching
fidelities exceeding 90\%,  even in
the presence of charge noise at the level of $2\mu$eV~\cite{Benito2019}. Since the fidelity is limited by the cooperativity $C=g_{\sigma}^2/\gamma_{\sigma}\kappa$, improvements are possible via increasing the double QD-resonator coupling $g_c$ or reducing the spin qubit and/or photon decoherence rates $\gamma_{\sigma}$ and $\kappa$.
To increase the coupling rate, superconducting cavities with higher kinetic inductance that are to some extent resilient  to magnetic field are available~\cite{Samkharadze2016,Maleeva2018}.
Improvements in the photon decay rate are possible via  Purcell filters and improved resonator designs  if one relies on separate superconducting cavities for readout or gate-based readout~\cite{Rossi2017,Pakkiam2018,Urdampilleta2019,Zheng2019,West2019,Crippa2019}.
Reducing the spin qubit decoherence rate may be the most challenging, but one could try to reduce phonon emission~\cite{Rosen2019} and work at high-order sweet spots to reach a stronger protection against  charge noise~\cite{Benito2019b,AbadilloUriel2019}.

The use of hybrid architectures, embedding semiconductor  qubits in superconducting cavities, could potentially be an issue concerning the miniaturization and scaling. In this context, to truly harness the small size of the semiconductor qubits, one 
could benefit from the recent advances in the fabrication of QD arrays to increase the size of the computing nodes~\cite{Zajac2016,Uditendu2018,Mortemousque2018,Kandel2019,Volk2019,Lawrie2020,Chanrion2020,Ansaloni2020}. To this end, it is important to investigate short-distance coupling between spin qubits that have been proven to couple to cavity photons~\cite{Cayao2020}. Moreover, QD arrays allow for the exploration of new proposed qubits that couple to cavities but are more protected from decoherence such as the quadrupolar exchange only spin qubit \cite{Russ2018}.

 Eventually, for optimally controlled operations and for large-scale devices based on silicon it will be necessary to have a microscopic understanding and control of the valley features~\cite{Goswami2007,Yang2013}. Alternatively, some researchers are considering a shift from the conduction to the valence band, since  holes in silicon and germanium reside in a single non-degenerate valley~\cite{Maurand2016,Watzinger2018,Hendrickx2020}. Interestingly, holes also have a relatively strong spin-orbit interaction, which is particularly pronounced in germanium, an effect that could substitute the external micromagnets~\cite{Bulaev2005,Bulaev2007,Kloeffel2013b}.
 
Recent works have explored further superconducting-semiconducting hybrid quantum systems containing also superconducting qubits. They employ a general circuit QED architecture to demonstrate a coherent interface between semiconductor and superconducting qubits~\cite{Scarlino2019,Landig2019}.
In the future, superconducting cavities may on the one hand act as connectors between like qubits, and on the other hand  bridge between vastly different quantum systems.


\begin{acknowledgments}
\textit{Acknowledgments.---}
This work was supported by ARO through Grant No.~W911NF-15-1-0149
and the DFG through the collaborative research center SFB~767. 
\end{acknowledgments}

\textit{Data availability.---}
Data sharing is not applicable to this article as no new data were created or analyzed in this study.


\bibliographystyle{apsrev4-1}

\bibliography{postdocreferences.bib}

\end{document}